\pdfoutput=1
\documentclass[]{article}

\usepackage{cite}

\usepackage[pdftex]{graphicx}

\usepackage{array}

\usepackage{url}
\usepackage{graphicx}

\begin{document}
\title{\#Sleep\_as\_Android: Feasibility of Using Sleep Logs on Twitter for Sleep Studies\thanks{This is a preprint of an article accepted to appear at IEEE ICHI 2016.} }

\author{
Fatema Akbar
\qquad 
Ingmar Weber\\
\\
Qatar Computing Research Institute, HBKU\\
Doha, Qatar\\
{[}fakbar, iweber]@qf.org.qa
}

\date{}

\maketitle

\begin{abstract}
Social media enjoys a growing popularity as a platform to seek and share personal health information. For sleep studies using data from social media, most researchers focused on inferring sleep-related artifacts from self-reported anecdotal pointers to sleep patterns or issues such as insomnia. The data shared by ``quantified-selfers" on social media presents an opportunity to study more quantitative and objective measures of sleep. We propose and validate the approach of collecting and analyzing sleep logs that are generated and shared through a sleep-tracking mobile application. We highlight the value of this data by combining it with users' social media data. The results provide a validation of using social media for sleep studies as the collected sleep data is aligned with sleep data from other sources. The results of combining social media data with sleep data provide preliminary evidence that higher social media activity is associated with lower sleep duration and quality.
\end{abstract}

\section{Introduction}
People spend about 30\% of their life sleeping \cite{timesurvey}. Apart from being an integral part of a person's life, sleep also has an important influence on a person's mental and physical health \cite{spiegel2009effects, miller2007inflammation, cappuccio2010quantity, gangwisch2006short}, general well-being \cite{sleepwellbeing}, mood \cite{sleepmood} and productivity \cite{rosekind2010cost}. Despite the importance of sleep, population-level sleep analyses largely depend on self-reported sleep or observe a limited number of individuals in sleep laboratories. Although there have been recent attempts by researchers to use social media as a source of personal health information, these studies still use self-reported qualitative sleep measures, such as mentions of ``I can't sleep'' or ``insomnia'' by users on social media \cite{mciver2015characterizing, jamison2012can}. While those studies offer a valuable lens to look at sleep from the public health level, more details and more accurate measurements are needed for a closer investigation of sleep and well-being at the individual level.

Advances in wearable technologies and mobile applications, and the emerging phenomena of ``the quantified-self'' \cite{swan2009emerging}, allow for more accurate quantitative measures of sleep duration and quality. Several of these wearable technologies and mobile applications automatically generate sleep reports that can be shared on social media. This presents an opportunity to address a gap in previous studies that used social media as a source of information on sleep. When paired with other personal data from social media, the automatically generated sleep reports can potentially provide insights into the relationship between sleep and general aspects of an individual's well-being such as their social life on a personal-level, and at a large-scale. The combination of quantified self and social media data is an area that is promising and still largely unexplored \cite{wangetal06dh,HaddadiOMWS15}. Here, we present results on using sleep logs from social media for sleep studies, illustrating the value of this approach.

Concretely, our main contributions are as follows:

\begin{itemize}
\item We show how a combination of quantified self and social media data can be obtained to study sleep.
\item We describe biases and limitations of the collected data in terms of observed sleep patterns.
\item We show evidence that social media activity shortly before sleeping has a negative effect on the sleep quality.
\item We categorize users based on their social media activity and show that different groups are associated with different sleep patterns.
\end{itemize}

We hope that our work adds to the growing literature on quantified-self data by demonstrating the value of \emph{combining} this data with social media data at the individual level, therefore offering both physical and social sensing.

\section{Background}

\subsection{Social Media as a Data Source for Sleep Studies}
With the emerging concept of the ``Quantified-Self", sharing personal health information on social media has become common \cite{swan2009emerging}. Personal health information disclosed on social media by quantified selfers range from physical activity data, calories consumed, heart rate, to sleep-related data. Anecdotal personal health information, such as describing adverse drug reaction or discussing symptoms, is also shared by users on social media \cite{de2014seeking} and has been found to be promising for public health studies \cite{yang2012detecting, yang2013harnessing}. In sleep studies using social media as a data source, several approaches to collecting and analyzing data can be found. Some studies on sleep data from social media used anecdotal information provided by users. For example, \cite{mciver2015characterizing} and \cite{jamison2012can} collected tweets containing keywords such as ``insomnia" and ``can't sleep" and performed a linguistic content analysis as well as sentiment analysis to investigate how users discuss their sleep issues on social media. They both found that users sharing their sleep problems on social media have lower sentiment. McIver et al.~\cite{mciver2015characterizing} added data on social media activity of the users who tweeted a sleep-related keyword to compare the number of tweets and friends between users who disclose their sleep issues on Twitter and those who do not. They found that users with sleep problems have less social connections and are more active during typical sleep times. Wu et al.~\cite{wuetal15apweb} estimated sleep start and end times, as well as sleep quality, by analyzing microblogs from a microblogging platform in China and reported that a reasonable model can be designed to study sleep from social media data.

As the previous examples of related work show, social media can be a valuable source of data for sleep studies. However, most of the research in this area either estimate sleep-related data from social media activity or use self-reported anecdotal pointers to sleep patterns and/or issues. The opportunity to study sleep from more objective and quantitative data shared on social media through sleep-tracking applications remains untapped.

Quantified selfers share data collected by wearable technologies and mobile applications, which provides an opportunity for quantitative studies. Their data, potentially, offers higher level of precision and reliability than the anecdotal information shared on social media. We study feasibility of using this data and highlight both opportunities and challenges when conducting sleep studies using sleep-tracker data from social media.

\subsection{Clinical Validity of Sleep Tracking Mobile Apps}
Wearable technologies and mobile applications that track and measure sleep use various approaches. A review of sleep screening applications for smartphones reported that one common approach for assessing sleep through phone sensors is actigraphy \cite{behar2013review}. Actigraphy involves sensing physical motion movements of the user as an indication of sleep stages. This approach is used either by having a user wear a wrist band (or smart watch) or by using the accelerometer in mobile phones placed next to the user when he/she is asleep to sense motion. Another sleep sensing approach using smartphones, as noted by \cite{behar2013review}, is via audio recordings using the internal microphone of a mobile phone or an external one. The exact algorithms of how sleep is detected in different apps are proprietary, which makes scientific validation by the academic community difficult \cite{ong2016overview}.

There are some studies that compared sleep tracking mobile applications to gold standard in sleep studies (PSG method) and found that mobile applications do not detect sleep stages like PSG does \cite{bhat2014there}. However, the mobile applications tested in these validation studies were able to reasonably detect sleep vs.~wake states \cite{bhat2014there, bianchi2015consumer}. As noted by \cite{van2011objective}, ``[c]ompared with laboratory-based, PSG agreement rates in healthy volunteers were relatively high, but actigraphic accuracy tends to diminish in people with a lower sleep quality, because quiet wakefulness is scored as sleep". As they do not accurately detect sleep stages or sleep disorders, sleep-tracking mobile applications can be useful for general sleep tracking for well-being rather than diagnosis or other clinical purposes \cite{kelly2012recent}. Researchers and practitioners agree that sleep tracking mobile apps at present do not qualify as medical devices, but they have a great promise to study sleep from a general wellness perspective rather than a medical perspective.

Our app of interest, Sleep As Android, has not been evaluated against medically accepted methods in the literature \cite{ong2016overview}, but its sleep tracking technology and methods used are similar to other mobile applications that were validated in the aforementioned studies.

\subsection{`Sleep As Android' Mobile Application}

There are several applications for sleep tracking in the Android and Apple markets. In an overview of smartphone applications for sleep analysis, \cite{ong2016overview} reported that the most reviewed app by far is the Sleep as Android app. According to the app store's statistics in March 2016, the app was downloaded 10,000,000 - 50,000,000 times and had 202,021 user reviews. The app offers manual or automatic sharing of sleep logs on social media. Being a popular app and having the option to share logged sleep data on social media, Sleep As Android was considered a suitable mobile application for this study.

Users can download the mobile application from the Google Play Store (the store for Android Apps) and can record their sleep by instructing the app indicating when they go to bed and turning the tracking off when they wake up. In recording sleep start and end times, Sleep As Android's approach can be compared to sleep diaries where individuals manually log their sleep. As for sleep quality, indicated as ``deep sleep" by the app, the app's developers state that they use actigraphy (accelerometric movement measurement) to recognize sleep phases as more movements indicate lighter sleep. However, the algorithms to calculate deep sleep based on accelerometer data are not published. 

According to the app's webpage, the app can be synchronized with a smartwatch to measure heart rates and improve sleep tracking. It is also indicated that the sleep tracking works well with different mattress types excluding 100\% slow foam mattresses\footnote{\url{http://sleep.urbandroid.org/documentation/core/sleep-tracking/}}. Notably, measurements from Sleep As Andoid app have been used as a ground truth in an experiment to test a new sleep tracking app \cite{bai2012will}. To this end, we consider Sleep As Android as a suitable source for sleep data for this study. The limitations and caution needed in interpreting the results (e.g.~for medical studies) are discussed further in the discussion section.

\section{Methodology}
\subsection{Data Collection}
We have used several methods to collect tweets containing the hashtag \#Sleep\_as\_Android which is the hashtag included in the auto-generated tweets from the Sleep as Android sleep tracking mobile application.\footnote{See \url{https://twitter.com/search?f=tweets&q=\%23sleep_as_android&src=typd} for an example of such tweets.} To collect historic tweets, we used Twitter's search API which provides a short history of tweets. To go further back in Twitter's tweets history to search for relevant tweets, we used Topsy.com\footnote{Topsy.com was shut down in December 2015.} which indexes the entire Twitter timeline. 

\begin{figure}[!t]
\centering
\includegraphics[width=2.4in]{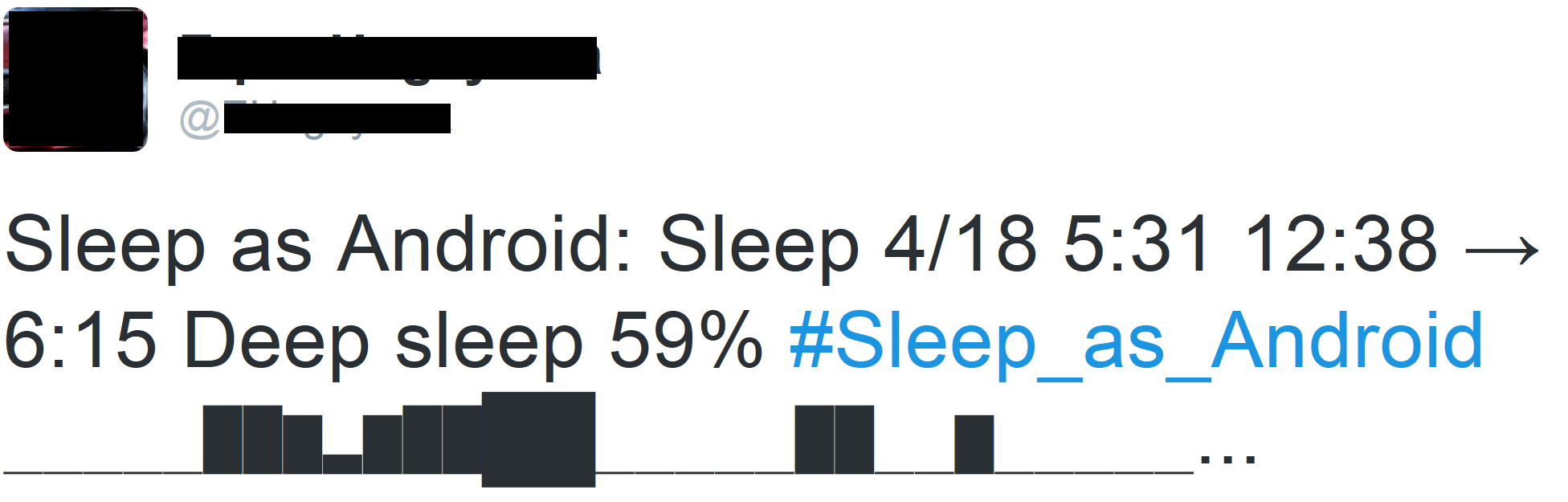}

\includegraphics[width=2.5in]{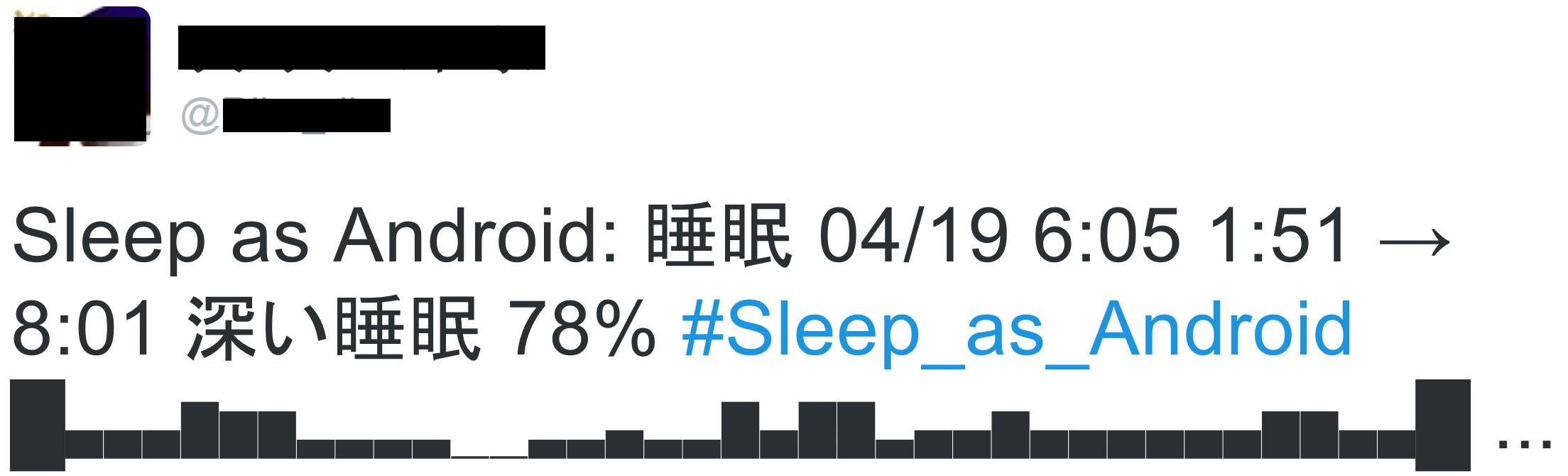}
\caption{Examples of valid auto-generated sleep-log tweets included in our dataset.}
\label{fig_sim}
\end{figure}

Besides collecting historic tweets, new tweets were collected by using Twitter's streaming API to capture any new tweet containing the relevant hashtag. The streaming job was running for a period of two months (October to November, 2015). Figure~\ref{fig_sim} shows some examples of the auto-generated sleep log tweets from the Sleep as Android mobile application.

\subsection{Data Cleaning and Pre-Processing}
Duplicate tweets were removed from the dataset. The tweet texts were parsed to extract sleep start time, sleep end time, sleep duration and sleep quality. Meta-data of the tweets were also parsed to obtain the user ID, tweet ID, tweet time-stamp and user data such as location, timezone, Twitter interface language and biography. To get users' countries, their timezone and self-stated location were collected from the meta-data of the tweets and a reverse geo-coding tool\footnote{\url{https://nominatim.openstreetmap.org/}} was queried to get the countries associated with these timezones and locations. In cases where there was no location or timezone, we use the user's Twitter interface language (which is included in their profile information) as a proxy for location.

\begin{figure}[!t]
\centering
\includegraphics[width =3in]{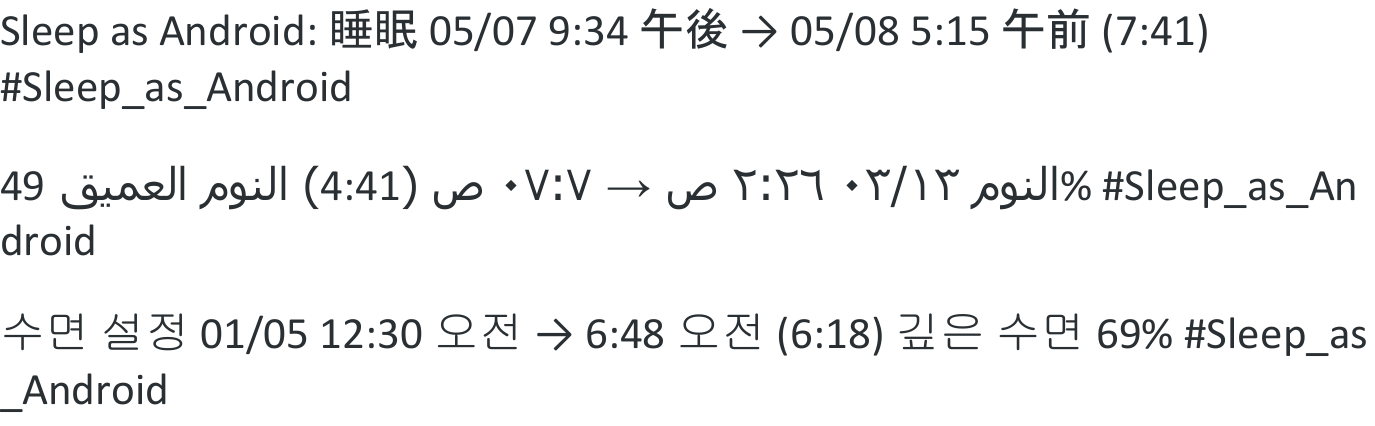}
\caption{Examples of tweets using non-English numbers and notations. Similar tweets are filtered-out from our dataset.}
\label{non_eng_twt}
\end{figure}

Sleep log tweets that use non-English numbers characters and/or non-English AM/PM notations are removed. Figure~\ref{non_eng_twt} shows some examples of tweets using non-English numbers and notations. However, tweets that are not in English but use English numerals and notations are included (e.g.~the second tweet in Figure~\ref{fig_sim}). Different notations for time were considered. Specifically, whether the time was reported as 12-hour or 24-hour format, whether the 12-hour format notation is AM/PM or a.m./p.m., and whether the hour/minute separator is a colon (e.g.~02:10) or a dot (e.g.~02.10) were all taken into consideration in the text mining. Tweets containing the hashtag but not following the standard pattern for the auto-generated tweet starting with ``Sleep as Android: " and not containing a sleep log were removed as these tweets are mostly advertisements, spam, or tweets of users using the hashtag to express their opinion about the app. Sleep logs with a sleep duration of less than 2 hours were removed as they may be from users who are testing the app or recording a nap. Sleep logs with a sleep duration of more than 12 hours were also removed as they may be from users who forgot to set the app properly to stop recording their sleep when they wake up and kept the app running.

\section{Results}
\subsection{Descriptive Statistics}
\begin{figure}[!t]
\centering
\includegraphics[width =2.5in]{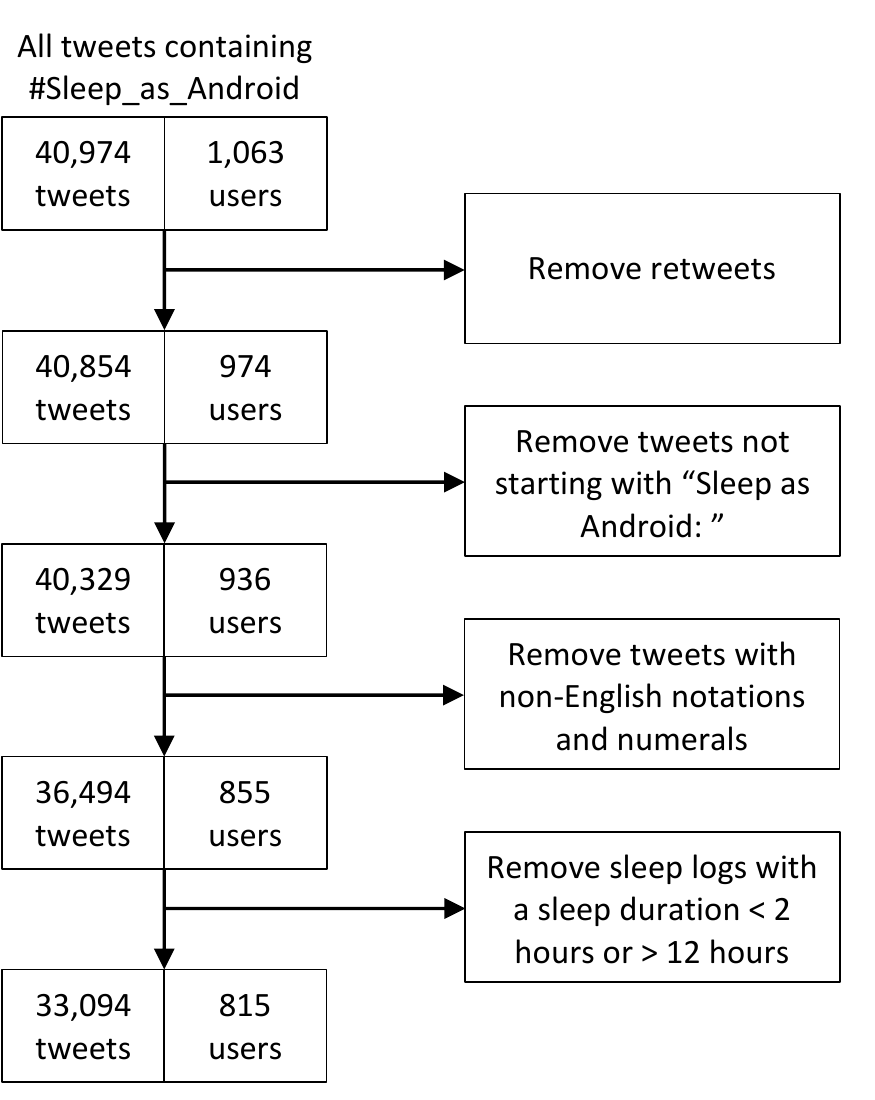}
\caption{The number of tweets and users throughout the filtering stages.}
\label{tweets_filtering_process}
\end{figure}

We have collected 40,974 tweets containing the hashtag \#Sleep\_as\_Android, of which 40,329 were valid sleep logs of 936 users. After filtering out tweets with non-English notations and numbers and sleep logs with a duration of less than 2 hours or more than 12 hours, 33,094 sleep logs of 815 users remained. Figure~\ref{tweets_filtering_process} shows the number of tweets and users throughout the filtering process. Table~\ref{freq_table} shows the number of sleep logs per user. As can be seen from Table~\ref{freq_table}, the majority of users tweet more than one sleep event.

\begin{table}[!t]

\renewcommand{\arraystretch}{1.3}

\caption{Frequency table of the number of sleep logs per user (Groups of the number of sleep logs are spaced on a logarithmic scale).}
\label{freq_table}
\centering

\begin{tabular}{|c|c|c|}
\hline
Number of sleep logs & Number of users & Percentage of users\\
\hline
1 & 194 & 24\%\\
\hline
2-3 & 129 & 16\%\\
\hline
4-7 & 91 & 11\%\\
\hline
8-15 & 104 & 13\%\\
\hline
16-31 & 92 & 11\%\\
\hline
32-63 & 64 & 8\%\\
\hline
64-127 & 66 & 8\%\\
\hline
128-255 & 46 & 6\%\\
\hline
256-633 & 29 & 3\%\\
\hline
Total & 815 & 100\%\\
\hline

\end{tabular}
\end{table}

\begin{figure}[!t]
\centering
\includegraphics[width =3in]{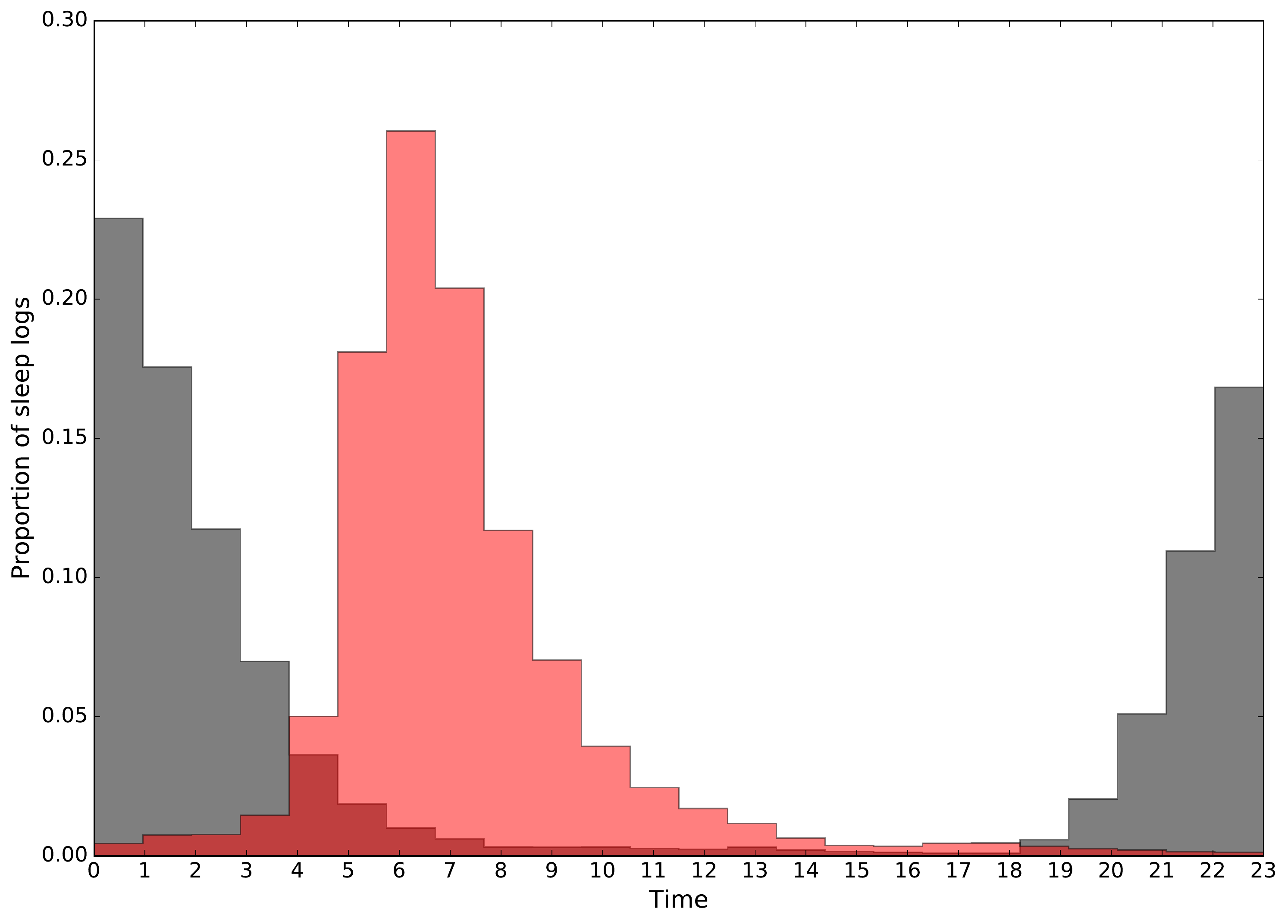}
\caption{The proportion of sleep logs per sleep start time (red) and sleep end time (grey).}
\label{sleep_start_and_end_hist_normed}
\end{figure}

The sleep logs showed that the most common time to go to sleep is between 10 PM and 3 AM (77\% of logs) as shown in Figure~\ref{sleep_start_and_end_hist_normed}. 25\% of the sleep logs showed a wake-up time between 6 to 7 AM, with the majority (80\%) of the logs showing wake-up times between 5 to 10 AM. Breaking the results by country, we found that users in Japan tend to sleep later than users in the U.S., but the distribution of wake-up times for both groups of users are similar with most users waking up between 5 to 7 AM.

The average sleep duration in the sleep logs is 6 hours (overall average is 6.25 hours and average of users' averages is 6.06 hours) with 53\% of users having an average sleep duration between 5 and 7 hours. To ensure that one-time users do not influence the results, we calculated these statistics on users with 5 or more sleep logs tweeted, and the distribution remained the same.

Examining the profiles of the users who use the sleep-tracking application and share their sleep logs on Twitter, we found users from 22 different timezones in 77 different cities. Around 44\% of the users are in Japan (either by their timezone selected, self-stated location, or interface language used), 14\% are in the U.S., 7\% in Russia and 4\% in the U.K.

\begin{figure}[!t]
\centering
\includegraphics[width =3in]{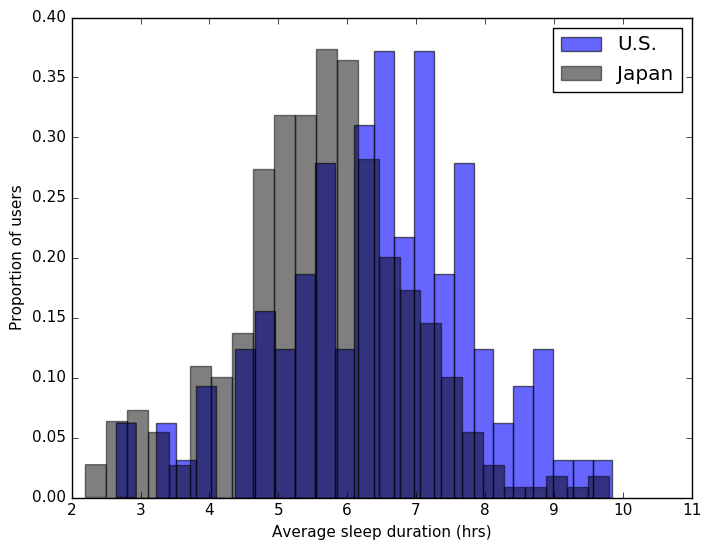}
\caption{Distribution of average sleep duration per user in the U.S. and Japan.}
\label{jp_us_sleep_d}
\end{figure}

\begin{figure}[!t]
\centering
\includegraphics[width =3in]{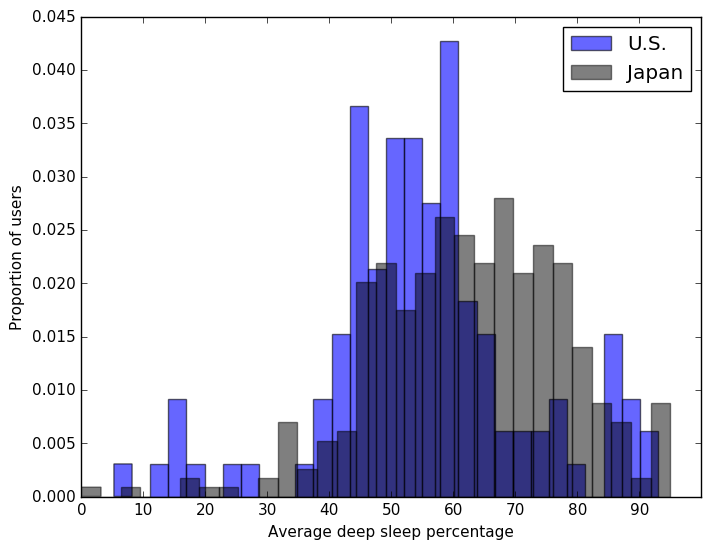}
\caption{Distribution of average deep sleep percentage per user in the U.S. and Japan.}
\label{jp_us_sleep_q}
\end{figure}

Given the large proportion of users from Japan and the U.S., we compared their distributions of the average sleep duration per user (Figure~\ref{jp_us_sleep_d}). The difference in the distribution of average sleep duration between the U.S. and Japan is statistically significant (Mann-Whitney U test p-value\textless 0.001). Japanese users have a lower average sleep duration than American users, and lower than all non-Japanese users. As for deep sleep, Japanese users get a higher deep sleep percentage than Americans or the rest of the countries (Figure~\ref{jp_us_sleep_q}).

\subsection{Sleep Time and Sleep Duration}
\begin{figure}[!t]
\centering
\includegraphics[width =3.5in]{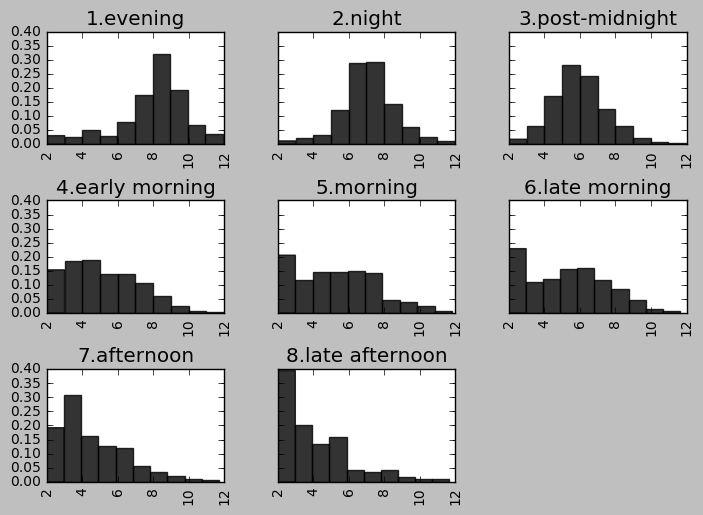}
\caption{Histograms for sleep duration by the time of the day sleep started.}
\label{Sleep_duration_hist_by_sleep_start_time_of_day}
\end{figure}

Notably, users who started their sleep between 7 PM and 9 PM had the longest sleep duration (8 hours on average) and the highest deep sleep percentage (73\% on average). A Mann–Whitney U test shows that the difference in sleep duration and deep sleep percentage between sleep starting in the evening (18:00 to 21:00) and sleep starting post-midnight (00:00 to 03:00) is statistically significant (p\textless 0.001 for both sleep duration and deep sleep tests). Figure~\ref{Sleep_duration_hist_by_sleep_start_time_of_day} shows a histogram matrix for sleep duration by the time of the day sleep started.

\subsection{Sleep Patterns Per Day of the Week}
\begin{figure}[!t]
\centering
\includegraphics[width =2.5in, trim={.8cm 8.3cm 1 0.8cm}]{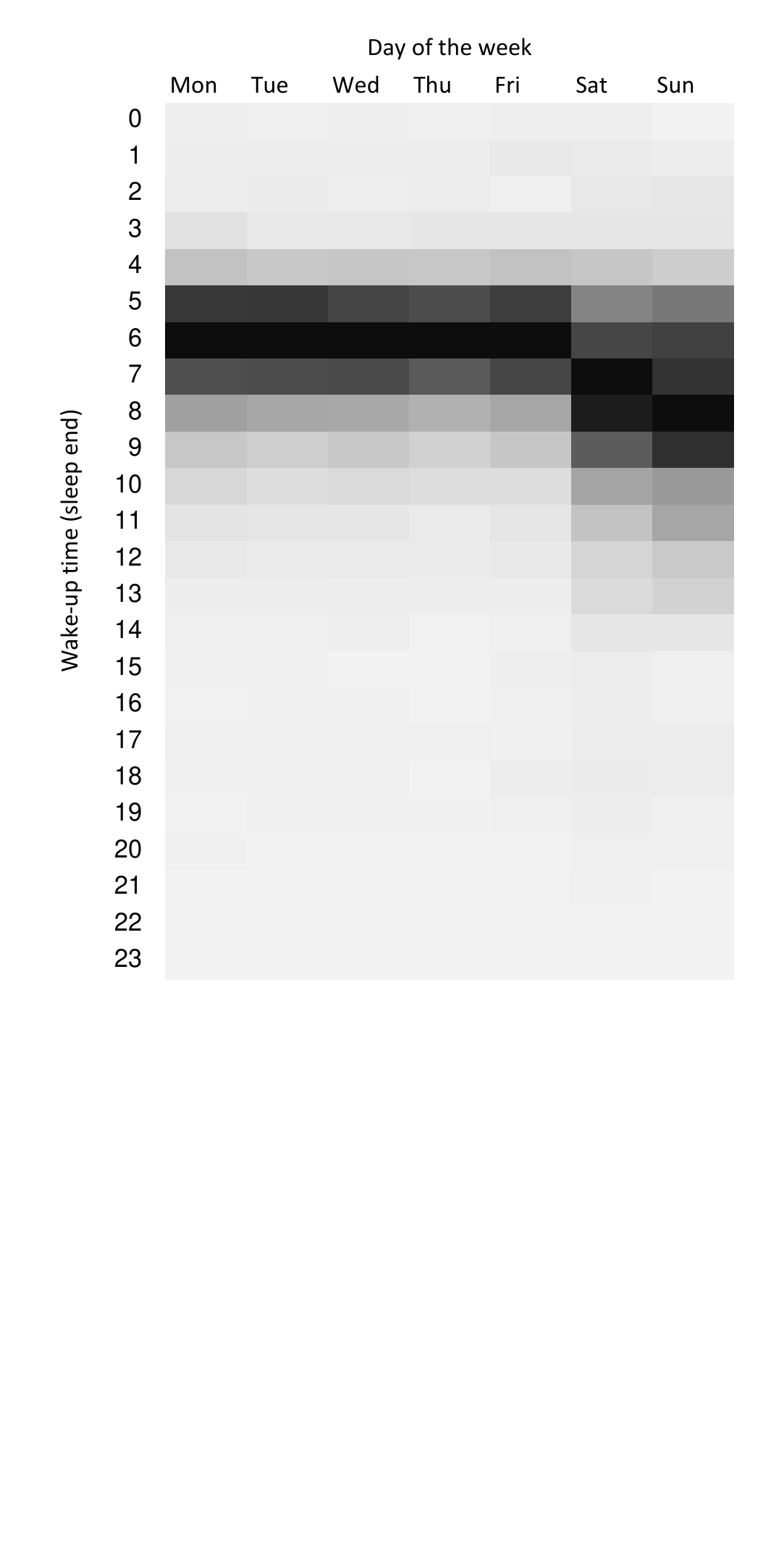}
\caption{A heatmap of the wake-up times for each day of the week.}
\label{sleepEnd_day_of_week_heatmap}
\end{figure}

The sleep logs clearly show that people tend to go to sleep later on weekends, and tend to wake-up later as well (Figure~\ref{sleepEnd_day_of_week_heatmap}). Deep sleep percentage does not seem to be different on weekends. 

\subsection{Combining Social Media Data with Sleep Data}

Manually inspecting some profiles to understand whether sleep patterns could be related to personal characteristics and lifestyle as seen through the activity and content seen on users' Twitter page, we found that some users create Twitter accounts mostly dedicated to tweeting their ``quantified-self" data from wearables and apps. Other users had regular Twitter accounts tweeting about their interests and daily activities, interacting with other Twitter users, and interacting with content such as news and media, which could be a rich source of data to study in relation to sleep patterns. 

As a step towards investigating how social media activity can be associated with sleep patterns, we correlated the tendency of users to tweet in the two-hour window prior to going to sleep time and their average sleep quality, and we found a weak negative correlation (r=-0.083, p=0.03). Dividing users into groups based on quartiles of their probability of tweeting before sleeping, we found a statistically significant difference in average sleep quality of users in the top quartile (i.e. users with highest tendency to tweet before sleeping) and users in the lower quartile (d=0.034, p-value=0.026).

\begin{figure}[!t]
\centering
\includegraphics[width =3.5in, trim={0.1cm 0.5cm 0 0.3cm}]{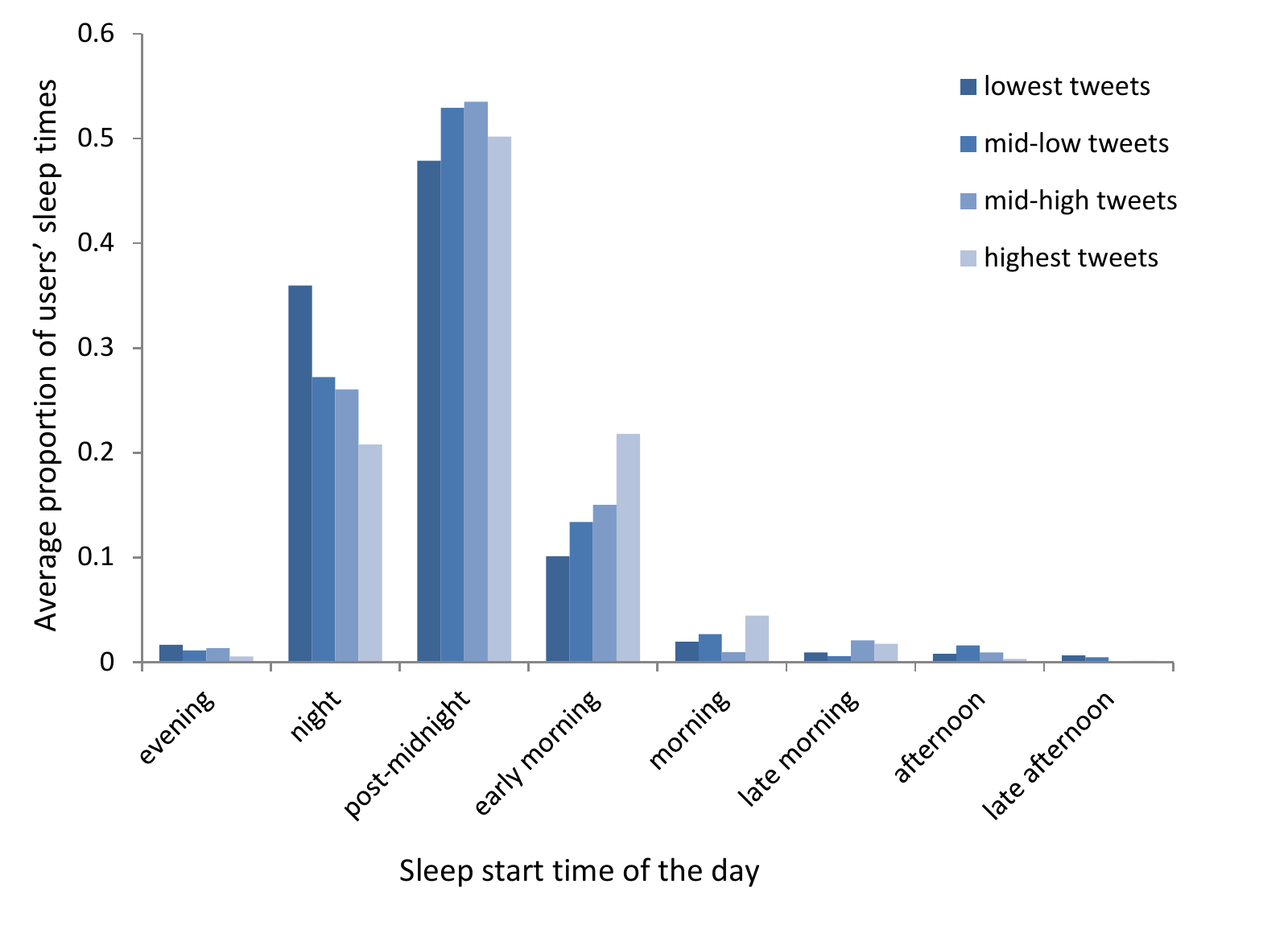}
\caption{Time of the day users tend to go to sleep based on how active they are on social media (measured by the number of tweets per day).}
\label{sleep_time_per_tweets}
\end{figure}

Aligned with the previous finding, we found a statistically significant difference (p\textless 0.001) in sleep duration between active tweeters and less active tweeters. We defined active and less active tweeters based on their average tweets per day. Dividing the users into four groups based on which quartile their number of average tweets per day fall into, we plotted the proportions of users' sleep logs that indicate waking up in each time of the day (Figure \ref{sleep_time_per_tweets}). Notably, less active tweeters tend to go to sleep at night (between 9 PM to midnight) more than more active tweeters. Active tweeters, on the other hand, tend to go to sleep in the early morning hours (3 to 6 AM) more than less active tweeters. We ran the same analysis for Japanese user and American users separately, and the same observations were noted, except that American users in general, as noted before, sleep earlier than Japanese users.

Additionally, dividing users based on the number of friends (followees) they have on Twitter, we found a statistically significant difference in sleep duration between users with more friends and users with less friends on Twitter (Mann-Whitney U test p-value=0.03).

Overall, combining social media data with sleep data can provide an insight on how activity on social media is related to different sleep patterns. Our findings show evidence that different levels of social media activity are related to different sleep duration, quality, and times. Further work in this area can reveal more about sleep patterns as they relate to different social networks patterns and the type of social media activity (e.g. using Twitter for information/news versus using Twitter for socializing). Directions for future research are discussed further in the conclusion.

\section{Discussion}

\subsection{Data Validation}
The sleep logs data we obtained from Twitter is aligned with sleep reports based on surveys and other wearable technologies. For example, Gallup researchers\footnote{\url{http://gallup.com/poll/166553/less-recommended-amount-sleep.aspx}} reported that Americans get 6.8 hours of sleep at night, which is similar to our finding of 6.46 hours. The different sleep pattern in the weekend is also reported by sleep survey results published by the Better Sleep Council\footnote{\url{http://bettersleep.org/better-sleep/the-science-of-sleep/sleep-statistics-research/starving-for-sleep-americas-hunger-games/}}.

Besides aligning with sleep statistics that are based on surveys, data collected directly by similar sleep-tracking devices, such as Sleep Cycle, Jawbone, and Withings, show similar results. For example, reports published by Sleep Cycle mobile application show a higher sleep quality but lower sleep duration for Japan compared to the U.S.\footnote{\url{http://www.sleepcycle.com/sleep-quality-world/}}, which was also found in our Sleep as Android data, as well as a similar finding on sleep times during weekdays and weekends\footnote{\url{http://www.sleepcycle.com/how-the-world-sleeps-days-of-the-week/}}. In addition, sleep reports published by Jawbone\footnote{\url{https://jawbone.com/blog/jawbone-up-data-by-city/}} and Withings\footnote{\url{http://blog.withings.com/2014/11/04/study-of-the-sleep-patterns/}} sleep trackers also align with our findings. Jawbone reports a sleep duration average of 5:44 hours for Japan (5:37 hours in our data) and 6:58 for Australia (6:40 hours in our data). Withings also reported the different sleep patterns on weekends.

Even though the sleep data we collected from social media align with data collected directly by other devices, it is important to note possible biases in the dataset. There might be inherent differences between people who share their quantified self data on social media and the rest of the population. The data might be biased towards ``healthy" or ``normal" sleepers as the majority of the sleep logs show sleep time within the typical sleep/wake periods.

Furthermore, sleep quality -or deep sleep as referred to here- should be interpreted with caution as many health researchers indicated the possible overestimation of deep sleep by mobile phone sensors and the weak correlation between deep sleep measured by mobile applications compared to clinical devices. 

Nevertheless, for general well-being studies, collecting, cleaning and analyzing sleep data from social media can provide a rich dataset covering a large sample of users which is often difficult in clinical settings. Using sleep data shared on social media is a feasible and valid approach to study sleep trends and patterns at the individual level or the public level.

\subsection{Sleep and Social Media Activity}

The advantage of using social media as a data source for sleep studies, besides having access to valid and reliable data for a large sample, is the access to other data that provides social sensing to complement the physical sensing of the mobile sleep tracker. Insight into personality and lifestyle can be gained from a user's activity on social media. The level of social media activity is also a valuable lens to examine sleep as it relates to how and why users use social media. Our preliminary findings in this direction of research provide important observations which highlight the value of combining social media data with sleep data.

Our findings relating high social media activity level with lower sleep duration and quality is aligned with other studies associating social media dependence to poor sleep quality by surveying users (e.g.~\cite{wolniczaketal13pone}), which again shows the feasibility and validity of using sleep data from social media and combining it with social media data of the users.

\subsection{Limitations}
Three key points could hinder the direct use of this data in public health studies:\\
1. Varying precision and accuracy of ``sleep quality" given the varying devices and sleep environments. \\
2. Intentional or unintentional errors by the users (e.g.~forgetting to stop sleep logging). Even after filtering out sleep logs with sleep duration \textless 2 hours or \textgreater 12 hours, there is still a possibility of errors in sleep logging by the users. This can affect the accuracy of the sleep duration measure or sleep start and end times.\\
3. Potential bias in the user base compared to the general public. ``Quantified-selfers" might differ from the public. Moreover, users might selectively share their data (e.g.~tweeting only good-sleep nights and deleting others). 

While the data gives a valid ``big picture", these limitations are especially important to consider when working with smaller subsets of the data, as more variation due to errors might distort the data.

\section{Conclusion}
This paper presented an approach to collect, clean and analyze sleep data from sleep logs generated by a sleep tracking mobile application and shared on Twitter by users of the app. Implementation and validation work have been conducted to assess the feasibility of this approach to sleep and social media studies. Combining sleep logs with social media data such as user profiles and tweets, we found evidence that different social media activity levels are associated with different sleep patterns. Specifically, we found that higher social media activity levels are associated with lower sleep quality and duration. These insights provide a strong motivation for future research combining the quantified-self data with other personal data from social media. Future work can apply topic modeling and sentiment analysis on users' social media content to investigate links of user interests, personality and lifestyle with their sleep data. Social networks of users can also be studied to investigate whether the types of social media connections (friends/followers) are related to a user's sleep pattern. Sleep-tracking mobile applications provide valuable data which can provide insight into the relationship between sleep and several aspects of an individual's well-being such as their social life on a personal-level and at a large-scale when combined with social media data.

\section*{Acknowledgment}
The authors would like to thank Dr.\ Shahrad Taheri and Dr.\ Teresa Arora from Weill Cornell Medicine in Qatar who provided insight and expertise on sleep studies.

\bibliographystyle{IEEEtran}
\bibliography{sleep-as-android}

\end{document}